\documentclass[aps,prl,twocolumn,superscriptaddress,showpacs,amssymb]{revtex4-1}
\usepackage{graphicx}
\usepackage{mathptmx}      % use Times fonts if available on your TeX system
\usepackage{subfigure}
\usepackage{color}

\newif\ifSHOWCOMMENTS
\newif\ifACCEPTCOMMENTS

\SHOWCOMMENTSfalse     % interchange to hide all comments
\SHOWCOMMENTStrue

\ACCEPTCOMMENTStrue    % Interchange to accept all replacements
\ACCEPTCOMMENTSfalse   %

\DeclareRobustCommand\openone{\leavevmode\hbox{\small1\normalsize\kern-.33em1}}

\begin{document}

\title{Quantum Decoherence at Finite Temperatures}

\author{M.A.\ Novotny}
\affiliation{Department of Physics and Astronomy, Mississippi State University,
Mississippi State, MS 39762-5167, USA}
\affiliation{HPC$^2$ Center for Computational Sciences, Mississippi State University,
Mississippi State, MS 39762-5167, USA}
\author{F. Jin}
\affiliation{Institute for Advanced Simulation, J\"ulich Supercomputing Centre,\\
Research Centre J\"ulich, D-52425 J\"ulich, Germany}
\author{S. Yuan}
\affiliation{Institute for Molecules and Materials, Radboud University of Nijmegen, \\
NL-6525AJ Nijmegen, The Netherlands}
\author{S. Miyashita}
\affiliation{
Department of Physics, Graduate School of Science,\\
University of Tokyo, Bunkyo-ku, Tokyo 113-0033, Japan
}
\affiliation{
CREST, JST, 4-1-8 Honcho Kawaguchi, Saitama, 332-0012, Japan}
\author{H. De Raedt}
\affiliation{Department of Applied Physics, Zernike Institute for Advanced Materials,\\
University of Groningen, Nijenborgh 4, NL-9747AG Groningen, The Netherlands}
\author{K. Michielsen}
\affiliation{Institute for Advanced Simulation, J\"ulich Supercomputing Centre,\\
Research Centre J\"ulich, D-52425 J\"ulich, Germany}
\affiliation{%
RWTH Aachen University, D-52056 Aachen,
Germany
}%

\date{\today}

\begin{abstract}
We study measures of decoherence and thermalization of a quantum system $S$ in the presence of a
quantum environment (bath) $E$. The whole system is prepared
in a canonical thermal state at a finite temperature.
Applying perturbation theory with respect to the system-environment coupling strength, we find that under
common Hamiltonian symmetries, up to first order in the coupling strength it is sufficient to
consider the uncoupled system
to predict decoherence and thermalization measures of $S$.
This decoupling allows closed form expressions for perturbative expansions for the
measures of decoherence and thermalization in terms of the free energies of $S$ and of $E$.
Numerical results for both coupled and decoupled systems with up to 40 quantum spins validate these findings.
\end{abstract}

\maketitle

Decoherence and thermalization are two basic concepts
in quantum statistical physics~\cite{KUBO85}. Decoherence renders
a quantum system classical due to the loss of phase coherence of the
components of a system in a quantum superposition via
interaction with an environment (or bath). Thermalization drives the system
to a stationary state, the (micro) canonical ensemble via energy exchange
with a thermal bath. As the evolution of a quantum system is governed by the
time-dependent Schr\"{o}dinger equation, it is natural to
raise the question how classicality could emerge from a pure quantum state.
This question is becoming more important technologically, for example in designing
quantum computers \cite{LADD2002} where decoherence effects are a major impediment in
engineering implementations.

Various theoretical and numerical studies have been performed,
trying to answer this fundamental question,
\textit{e.g.}, the microcanonical thermalization of an isolated quantum
system~\cite{YUKA11,NEUM29,PERE84,DEUT91}, canonical thermalization of a
system coupled to a (much) larger environment~\cite%
{TASA98,GOLD06,POPE06,REIM07,Yuan2009,YUKA11,LIND09,SHOR11,REIM10,GENW10,GENW13,GEMM06a}, and of two
identical quantum systems at different temperatures~\cite{PONO2011,PONO12}.
In our earlier work~\cite{JIN13a}, we found that at infinite temperature the degree of decoherence of a system $S$
scales with the dimension of the environment $E$ if
the state of the whole system $S+E$ is randomly chosen from the Hilbert space of the whole system.
We showed that in the thermodynamic limit, the system $S$ decoheres
thoroughly.

In this Letter, we investigate
for the first time quantitatively how classicality emerges from a pure quantum state when
$S+E$ is of a finite size and at a finite temperature.
We assume that the whole system $S+E$ is in a canonical thermal state,
a pure state at finite inverse temperature $\beta$~\cite{HAMS00, JIN10x, SUGI13},
and investigate measures of the decoherence and the thermalization of $S$.
This canonical thermal state could be the result of a thermalization process of the whole system $S+E$
coupled to a large heat bath, which we do not consider any further.
The state of the system $S$ is described by the reduced density matrix.
The degree of decoherence of the system $S$ is measured in terms of $\sigma$,
defined below in terms of measures of the off-diagonal components of the reduced density matrix.
If $\sigma=0$, then the system $S$ is in a state of full decoherence.
The difference between the diagonal elements of the reduced density matrix and the canonical or Gibbs
distribution is expressed by $\delta$. Hence, for a system being in its canonical distribution
it is expected that both $\sigma$ and $\delta$ are zero.
We show that under symmetry transformations that leave the
Hamiltonians of $S$ and $E$ invariant but
reverse the sign of the interaction Hamiltonian,
conditions which are usually satisfied for example in
quantum spin systems, the first-order term of the perturbation expansion of
$\sigma^2$ in terms of the interaction between $S$ and $E$ is exactly zero.
Therefore it is sufficient to study the uncoupled system.
We demonstrate that the leading term
in the expressions for $\sigma^2$ and $\delta^2$
is a product of factors of the free energy of $E$ and the free energy of $S$.
Hence, these expressions for $\sigma^2$ and $\delta^2$ allow one to study the influence
of the environment on the decoherence and thermalization of $S$.
We study the validity region of these results obtained from perturbation theory
by performing large-scale simulations for spin-$1/2$ systems with initial states from
a canonical thermal state.

The Hamiltonian of the
whole system $S+E$ can be expressed as
\begin{equation}
H=H_{S}+H_{E}+\lambda H_{SE}\>,  \label{Eq:H}
\end{equation}%
where $H_{S}$ and $H_{E}$ are the system and environment Hamiltonian,
respectively, and $\lambda H_{SE}$ describes the interaction between the
system $S$ and environment $E$. Here $\lambda$ denotes the global system-environment coupling
strength.

The state of the system $S$ is described by the reduced density matrix
$%
\hat{\rho}\equiv \mathbf{Tr}_{E}\rho \>,  \label{eq1}
$
where $\rho  =\left\vert \Psi \right\rangle \left\langle
\Psi \right\vert $ is the density matrix of the whole system $S+E$
and $\mathbf{Tr}_{E}$ denotes the trace over the degrees of freedom
of the environment $E$.
The state of the whole system $S+E$ can be written as $\left\vert \Psi
\right\rangle =\sum_{i=1}^{D_{S}}\sum_{p=1}^{D_{E}}c(i,p)\left\vert i,p\right\rangle \>,$
where the set of states $\{|i,p\rangle \}$ denotes a complete set of
orthonormal states in some chosen basis, and $D_{S}$ and $D_{E}$ are the
dimensions of the Hilbert spaces of the system $S$ and environment $E$,
respectively. We assume that $D_{S}$ and $D_{E}$ are both finite and $%
D=D_{S}D_{E}$ is the dimension of the Hilbert space of the whole system $S+E$%
. In terms of the expansion coefficients $c(i,p)$, the matrix element $%
(i,j)$ of the reduced density matrix reads $\hat{\rho}_{ij}=%
\sum_{p=1}^{D_{E}}c^{\ast }(i,p)c(j,p)\>.$

We characterize the degree of decoherence of the system $S$ by~\cite{Yuan2009}
\begin{equation}
\sigma =\sqrt{\sum_{i=1}^{D_{S}-1}\sum_{j=i+1}^{D_{S}}\left\vert
\widetilde{\rho }_{ij}\right\vert ^{2}}\>,  \label{eqsigma}
\end{equation}%
where $\widetilde{\rho }_{ij}$ is the matrix element $(i,j)$ of the
reduced density matrix $\hat{\rho}$ in the representation that diagonalizes $%
H_{S}$. Clearly, $\sigma $ is a global measure for the size of the
off-diagonal terms of $\widetilde{\rho }$. If $\sigma =0$ the system is
in a state of full decoherence (relative to the representation that
diagonalizes $H_{S}$).
We define a quantity measuring the
difference between the diagonal elements of $\widetilde{\rho }$ and
the canonical distribution as~\cite{Yuan2009}
\begin{equation}
\delta =\sqrt{\sum_{i=1}^{D_{S}}\left( \widetilde{\rho }_{ii}-\left. {%
e^{-bE_{i}^S}}\right/ {\sum_{i=1}^{D_{S}}e^{-b E_{i}^S}}%
\right) ^{2}},  \label{eqdelta}
\end{equation}%
where $\{E_i^S\}$ denote the eigenvalues of $H_S$ and $b$ is a fitting inverse temperature.
The quantities $\sigma $ and $%
\delta $ are respectively general measures for the decoherence and thermalization of $S$.

A canonical thermal state is a pure state at a finite inverse
temperature $\beta $ defined by (the imaginary-time projection)~\cite{HAMS00, JIN10x, SUGI13}
\begin{equation}
\left\vert \Psi (\beta )\right\rangle
=
\left. e^{-\beta H/2}\left\vert \Psi(0)\right\rangle
\right /
\left\langle \Psi (0)\right\vert e^{-\beta H}\left\vert \Psi (0)\right\rangle ^{1/2} \> ,
\label{psi_ft}
\end{equation}%
where $\left\vert \Psi (0)\right\rangle =\sum_{i=1}^{D}d_{i}\left\vert
i\right\rangle $ and $\{d_{i}\}$ are complex Gaussian random numbers.
The justification of this definition can be seen from the fact that for any quantum
observable $A$ of the whole system~\cite{HAMS00, SUGI13}, one has $\left\langle
\Psi (\beta )\right\vert A\left\vert \Psi (\beta )\right\rangle \approx
\mathbf{Tr}Ae^{-\beta H}/\mathbf{Tr}e^{-\beta H}.$
The error in the approximation is of the order of the inverse square root of the Hilbert
space size of the whole system~\cite{HAMS00} and therefore the approximation improves for increasing $D$.
One may consider the state $\left\vert \Psi (\beta
)\right\rangle $ as a \textquotedblleft typical" canonical thermal state~%
\cite{SUGI13} in the sense that if one measures observables, their expectation
values agree with those obtained from the canonical distribution at the
inverse temperature $\beta $.
In the following we present analytical
results for $\sigma $ and $\delta $ when the whole system $S+E$ is in such a
state $\left\vert \Psi (\beta )\right\rangle $.

The system $S$ is coupled to an environment $E$
with global coupling strength $\lambda$, and therefore we resort to
perturbation theory with respect to $\lambda$.
Up to first order in the global system-environment coupling strength $\lambda$, we have~\cite{WILC67}
\begin{equation}
e^{-\beta {H}} \approx \left(1 - \left\{ \int_0^1 d\xi e^{-\beta\xi%
{H}_0} {H}_{SE} e^{\beta\xi{H}_0}\>\> \right\} \beta
\lambda \right)e^{-\beta {H}_0},
\label{expbH}
\end{equation}
where ${H}_0=H_S+H_E$ denotes the Hamiltonian of the uncoupled system and bath.
If we now consider a unitary transformation that leaves $H_S$ and $H_E$ invariant but reverses the sign of
$H_{SE}$, then
$\left < \Psi(0)\right | e^{-\beta {H}}\left | \Psi(0)\right > \approx {\bf Tr} e^{-\beta H_0}/D =Z_0/D$,
where $Z_0$ denotes the partition function of the uncoupled system.
The expression Eq.~(\ref{psi_ft}) for the canonical thermal state becomes
\begin{equation}
\left|\Psi(\beta)\right\rangle
\>\approx \>
\sqrt{\frac{D}{Z_0}}\> e^{-\beta H/2} \left|\Psi(0)\right\rangle.
\label{psi_ft2}
\end{equation}
Using Eq.~(\ref{expbH}) and Eq.~(\ref{psi_ft2}), we find that the first-order term of the
perturbation expansion in $\lambda$ of the expectation value of $\sigma^2$ is given by
\begin{eqnarray}
&& \left\langle\left\langle\sigma^2\right\rangle\right\rangle_1
 =
 - \beta\lambda \left(\frac{D}{Z_0}\right)^2\frac{D}{D+1}    \cr
&&\times   \left[
 Z_S {\bf Tr} e^{-\beta H_S}e^{-2\beta H_E} H_{SE}
\right .
  \left . -
 {\bf Tr} e^{-2\beta H_0}H_{SE}
 \right ],
\label{firstorder}
\end{eqnarray}
where $Z_S={\bf Tr} e^{-\beta H_S}$.
Here and in the following $\left\langle\left\langle\cdot\right\rangle\right\rangle$ denotes
the expectation value with respect to the probability distribution
of the random numbers $\{d_i\}$.
Applying the same unitary transformation as discussed before results in $\left\langle\left\langle\sigma^2\right\rangle\right\rangle_1=0$.
Hence, to study the decoherence of a system $S$ coupled to an environment $E$
up to first order in $\lambda$ it is {\it sufficient to study the uncoupled system\/} ($\lambda=0$).
Furthermore, one state of the type Eq.~(\ref{psi_ft}) is enough to get the correct expectation value
provided the dimension of the Hilbert space is large~\cite{HAMS00}.

Therefore, we now focus on the uncoupled system.
There exist simple relations for the eigenvalues $E_{j}$ (eigenstates $%
|E_{j}\rangle $) of the Hamiltonian $H$ in terms of the eigenvalues $%
E_{i}^{S}$, $E_{p}^{E}$ (eigenstates $|E_{i}^{S}\rangle $, $%
|E_{p}^{E}\rangle $) of the system Hamiltonian $H_{S}$ and environment
Hamiltonian $H_{E}$, respectively, \textit{i.e.}, $E_{j}=E_{i}^{S}+E_{p}^{E}$ and $%
\left\vert E_{j}\right\rangle =\left\vert E_{i}^{S}\right\rangle \left\vert
E_{p}^{E}\right\rangle $.
We first perform a Taylor series expansion of $\sigma^2$ up to second order in $|d|^2$ about the point $1/D$ and then
calculate the expectation value of $\sigma^2$. A lengthy calculation gives
\begin{eqnarray}
&&\left\langle\left\langle\sigma^2\right\rangle\right\rangle= \frac{1}{2}e^{-2\beta \left(F_E(2\beta)-F_E(\beta)\right)}
  \left(1-e^{-2\beta (F_S(2\beta)-F_S(\beta))}\right)  \cr
&& - \frac{2D}{D+1}e^{-3\beta (F_E(3\beta)-F_E(\beta))}  \cr
&&\times \left(e^{-2\beta (F_S(2\beta)-F_S(\beta))}-e^{-3\beta (F_S(3\beta)-F_S(\beta))}  \right)   \cr
&& + \frac{3D}{2(D+1)}e^{-4\beta (F_E(2\beta)-F_E(\beta))}e^{-2\beta (F_S(2\beta)-F_S(\beta))}		\cr
&&\times \left(1-e^{-2\beta (F_S(2\beta)-F_S(\beta))} \right),
\label{sigma}
\end{eqnarray}
where $F_{S}(n\beta )$ and $F_{E}(n\beta )$ represent the free energy of the
system $S$ and the environment $E$ at the inverse temperature $n\beta $,
respectively. Obviously, in most of the cases the first term dominates.

Two special cases are of interest. First, if $\beta =0$, we
recover the previous result $E(\sigma ^{2})=(D_{S}-1)/2(D+1)$~\cite{JIN13a}. In the vicinity of $\beta =0$, the first-order term of the Taylor expansion of
Eq.~(\ref{sigma}) vanishes. Hence in the high temperature limit,
$\left\langle\left\langle\sigma^{2}\right\rangle\right\rangle=(D_{S}-1)/2(D+1)+\mathcal{O}(\beta ^{2})$.

If the temperature approaches zero, Eq.~(\ref{sigma}) becomes
\begin{equation}
\label{Eq:gg}
\lim_{\beta \rightarrow \infty }\left\langle\left\langle\sigma ^{2}\right\rangle\right\rangle=\frac{g_{S}-1}{2g_{S}g_{E}}%
\left( 1-\frac{D_SD_E}{\left(D_S D_E+1\right)g_{S}g_{E}}\right) ,  \label{sigmalt}
\end{equation}%
where $g_{S}$ and $g_E$ refer to the degeneracy of the ground state of the system $S
$ and environment $E$, respectively. This expression yields zero if the
ground state of the system is non-degenerate. For a system with a highly degenerate
ground state ($g_{S}\gg 1$) the expression goes to $1/2g_{E}$.
For a system with known $g_S>1$ and a large environment $D_E \gg 1$, at small $\lambda$
and at low temperatures, if one measures
$\left\langle\left\langle\sigma^2\right\rangle\right\rangle$, one can determine the degeneracy $g_E$ of the ground state
of the environment.  This is a new, strong prediction.   The ground state degeneracy $g_E$ of the environment can be esitmated
by only measuring quantities in $S$.  Furthermore, Eq.~(\ref{Eq:gg}) should be a guide to engineering quantum
systems that have limited decoherence at low temperatures, namely systems with large $g_{S}$
(one characteristic of topological quantum computation \cite{NAYAK2008}) but non-degenerate $g_{E}$ so
$\left\langle\left\langle\sigma^2\right\rangle\right\rangle\approx 1/2$.

Similarly, we can make the Taylor expansion for $\delta^2$ up to second order with respect to both $%
|d|^2$ and $b$ about the points $1/D$ and $\beta$. The expectation value
of $\delta^2$ is given by
\begin{eqnarray}
&& \left\langle\left\langle\delta^2\right\rangle\right\rangle=\frac{D}{D+1} e^{-2\beta (F_E(2\beta)-F_E(\beta))}
\left(e^{-2\beta (F_S(2\beta)-F_S(\beta))} \right . \cr &&\left.
-2e^{-3\beta (F_S(3\beta)-F_S(\beta))}+e^{-4\beta
(F_S(2\beta)-F_S(\beta))}\right) \cr &&+e^{-2\beta (F_S(2\beta)-F_S(\beta))}%
\left[C_S(2\beta)/(4k_{\rm B}\beta^2) \right. \cr && \left. +( U_S(2\beta)
 -U_S(\beta) )^2)\right] (\Delta b)^2,
\label{delta}
\end{eqnarray}
where $\Delta b=b-\beta$, $k_{\rm B}$ is the Boltzmann factor, $C_S(n\beta)$ and $%
U_S(n\beta)$ are, respectively, the specific heat and average energy of the
system $S$ at inverse temperature $n\beta$. It is obvious that for the
uncoupled system $b=\beta$. For the coupled system, $b$ is not
necessarily equal to $\beta$.

In order to test the perturbation theory predictions,
we numerically simulate a spin-$1/2$ system divided into a system $S$ and an environment $E$
in a canonical state.
We consider a quantum spin-$1/2$ model defined by the Hamiltonian of Eq.~(\ref{Eq:H}) where
\begin{eqnarray}  \label{hamiltonian}
H_{S} &=&-\sum_{i=1}^{N_{S}-1}\sum_{j=i+1}^{N_{S}}\sum_{\alpha
=x.y,z}J_{i,j}^{\alpha }S_{i}^{\alpha }S_{j}^{\alpha },  \label{HAMS} \\
H_{E} &=&-\sum_{i=1}^{N_E-1}\sum_{j=i+1}^{N_E}\sum_{\alpha =x,y,z}\Omega
_{i,j}^{\alpha }I_{i}^{\alpha }I_{j}^{\alpha },  \label{HAME} \\
H_{SE} &=&-\sum_{i=1}^{N_{S}}\sum_{j=1}^{N_E}\sum_{\alpha =x,y,z}\Delta
_{i,j}^{\alpha }S_{i}^{\alpha }I_{j}^{\alpha }.  \label{HAMSE}
\end{eqnarray}%
Here, $S_i^\alpha$ and $I_i^\alpha$ denote the spin-$1/2$ operators of the spins at site $i$ of the system
$S$ and the environment $E$, respectively. The number of spins in $S$ and $E$
are denoted by $N_S$ and $N_E$, respectively. The total number of spins in
the whole system is $N=N_S+N_E$.
The parameters $J_{i,j}^\alpha$ and $\Omega_{i,j}^\alpha$ denote the spin-spin interactions of
the system $S$ and environment $E$, respectively and $\Delta_{i,j}^\alpha$ denotes
the local coupling interactions between spins of $S$ and spins of $E$.

From Eqs.~(\ref{HAMS}-\ref{HAMSE}) it is clear that the Hamiltonian of the whole system, Eq.~(\ref{Eq:H}),
obeys the symmetry properties which are required to make the first-order term of the perturbation expansion
of the expectation value of $\sigma^2$ (see Eq.~(\ref{firstorder})) exactly zero.
Namely, by only reversing the spin components of the system or environment spins, $H_S$ and $H_E$ do not change
but the sign of $H_{SE}$ reverses. Note that such a symmetry is also obeyed in the case that there is no interaction
between the environment spins, \textit{e.g.} for an environment Hamiltonian
$H_{E}=-\sum_{i=1}^{N_{E}}\sum_{\alpha =x,y,z}\Omega
_{i}^{\alpha }I_{i}^{\alpha }$~\cite{NOVO12b,NOVO12a}.
In this particular case, it is only required that $H_S$ is an even function and $H_{SE}$ an odd function under
reversal of all spin components of the system spins.

In our simulations we use the spin-up -- spin-down basis and use units such
that $\hbar =1$ and $k_{\rm B}=1$ (hence, all quantities are dimensionless).
Numerically, the
propagation by $\exp(-\beta H)$ is carried out by means of exact diagonalization and
the Chebyshev polynomial algorithm~\cite{TALE84,LEFO91,IITA97,DOBR03,RAED06} with initial state $\left\vert \Psi
(0 )\right\rangle $ (see Eq.~(\ref{psi_ft})). These algorithms yield results that are very
accurate (close to machine precision).

\begin{figure}[t]
\includegraphics[width=8cm]{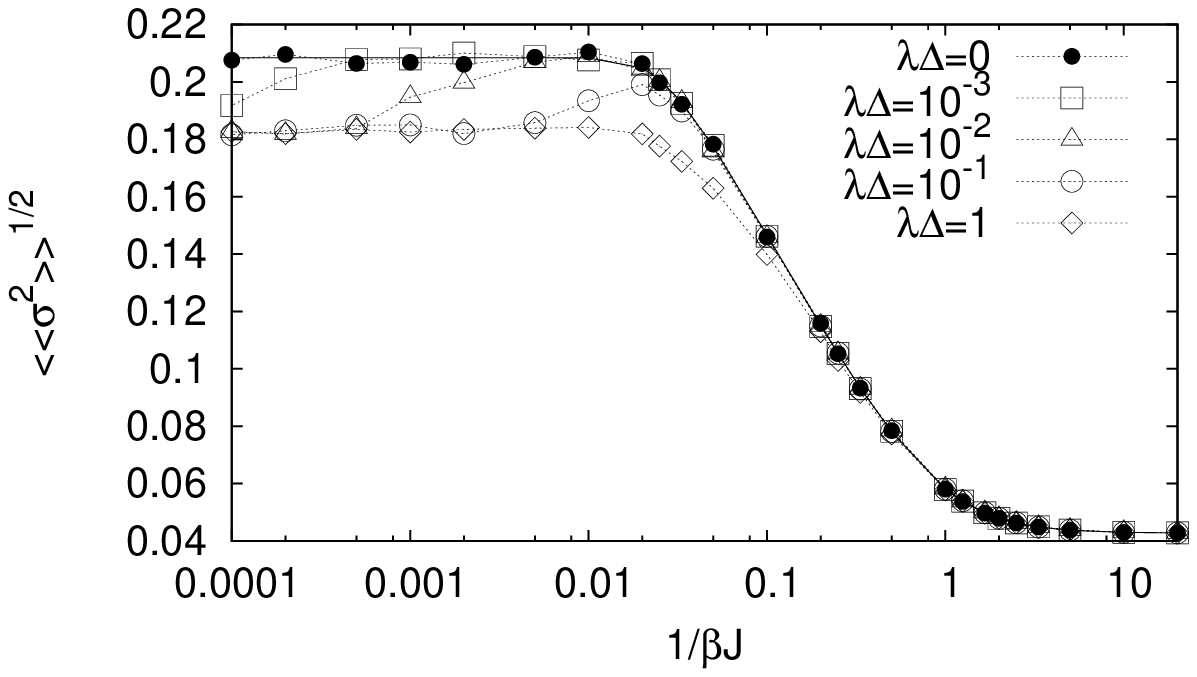} %
\includegraphics[width=8cm]{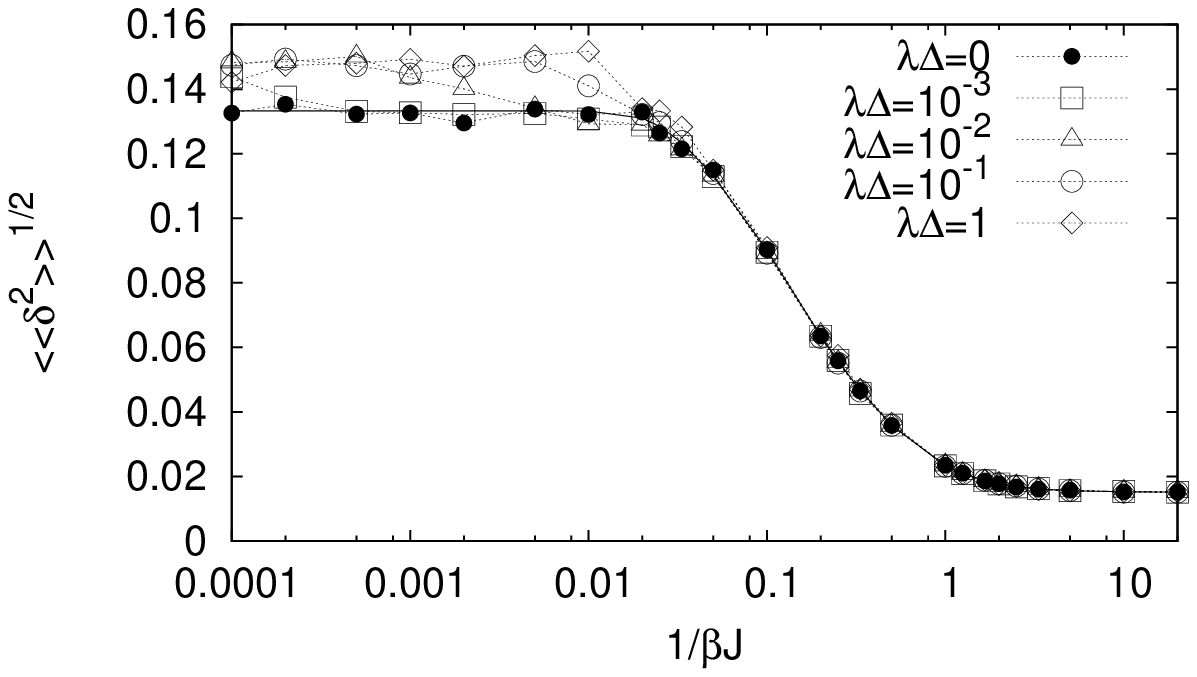}
\caption{Simulation results for ${\left\langle\left\langle\protect\sigma ^{2}\right\rangle\right\rangle}^{1/2}$ and $%
{\left\langle\left\langle\protect\delta ^{2}\right\rangle\right\rangle}^{1/2}$ for spin-$1/2$ chains with $N_{S}=4$,
$N_{E}=8$, $J=\Omega=1$ and various interaction strengths $\Delta$
as a function of the temperature $1/\beta J$. The solid lines are obtained from Eq.~(\ref{sigma}) and Eq.~(\ref{delta})
by using numerical values for the free energies $F_S(n\beta)$ and $F_E(n\beta)$.
The dotted lines are guides to the eye.}
\label{fig1}
\end{figure}

Figure~\ref{fig1} shows the simulation results for ${\left\langle\left\langle\sigma ^{2}\right\rangle\right\rangle}^{1/2}$
and ${\left\langle\left\langle\delta ^{2}\right\rangle\right\rangle}^{1/2}$ obtained by exact diagonalization
for the whole system $S+E$ being a spin chain with $N_{S}=4$ and $N_{E}=8$. The
system $S$ and environment $E$ consist of two ferromagnetic spin chains with
isotropic spin-spin interaction strengths $J_{i,j}^\alpha=J=\Omega_{i,j}^\alpha=\Omega =1$.
They are connected by one of their end-spins, with an interaction strength $\Delta_{N_S,1}^\alpha=\Delta$.
The global system-environment coupling strength $\lambda=1$.
The simulation results are averaged
over $1000$ runs with different initial random states.
Substituting the numerically obtained values for the free energy of the system and environment for $\lambda\Delta=0$
in the analytical expressions for $%
{\left\langle\left\langle\sigma^2\right\rangle\right\rangle}^{1/2} $ and ${\left\langle\left\langle\delta^2\right\rangle\right\rangle}^{1/2}$
given by Eq.~(\ref{sigma}) and Eq.~(\ref{delta}), respectively,
results in the solid lines depicted in Fig.~\ref{fig1}.
It is clear that the simulation
results for the uncoupled system ($\lambda\Delta =0$) and for the coupled cases when
$\left(\beta J\right)^{-1} \gtrsim 6 \lambda \Delta$
agree with the analytical results for the whole range of temperatures.
As the temperature decreases, the state of the whole system $S+E$ approaches the ground state,
$\left\langle\left\langle \sigma^2\right\rangle\right\rangle$ becomes constant, its numerical value being given by Eq.~(\ref{Eq:gg}).
For the case at hand, $g_S=5$, $g_E=9$, $D_S=16$ and $D_E=256$ and
Eq.~(\ref{Eq:gg}) yields $\left\langle\left\langle \sigma^2\right\rangle\right\rangle^{1/2}=0.21$, in excellent
agreement with the numerical data.
In the coupled case and for small $1/\beta J$, $\left\langle\left\langle \sigma^2\right\rangle\right\rangle^{1/2}$ develops a plateau
different from that of the uncoupled case.
The dependence of this plateau on $\beta$ or $\lambda\Delta$ is nontrivial,
requiring a detailed analysis of how the ground state of $S+E$ leads to the reduced density matrix  of $S$
(in the basis that diagonalizes $H_S$).
In this respect, the $\beta$ or $\lambda\Delta$ dependence of the data show in Fig.~\ref{fig1} are somewhat special because
the ferromagnetic ground state of the system does not depend on $\lambda\Delta$.

For the spin system under study with $\lambda\Delta\neq 0$, the first-order term of
the perturbation expansion of the expectation value of $\sigma^2$ in terms of $\beta \lambda\Delta$
is exactly zero. Hence, for a weakly coupled system ($\lambda\Delta$ small) deviations from the analytical
results Eq.~(\ref{sigma}) and Eq.~(\ref{delta}) obtained for the uncoupled system ($\lambda\Delta=0$),
are, as expected, seen {\it only\/} in the low temperature region.  The numerical results (symbols) in Fig.~1 are in
excellent agreement with the predicted results (solid lines) as long as $\beta\lambda\Delta$ is small.
For a finite $\beta\lambda\Delta$, the plateaus at low temperature may or may not be reached
and therefore the perturbation results are no longer applicable.

In order to study the behavior of $\sigma$ and $\delta$ as a function of the global coupling interaction strength $\lambda$,
we performed large-scale simulations
for a spin system configured as a ring with $N_S=4$ and $N_E=26,36$ at the inverse temperature $\beta |J|=6$.
This required working with a Hilbert space of size up to $D$$\approx$$1.1$$\times$$10^{12}$.
We assumed that the spin-spin interaction strengths of the system $S$ are isotropic, $J_{i,j}^\alpha=J=-1$, that
those of the environment $E$ are only nonzero for nearest neighbors and that only $\Delta_{N_S,1}^\alpha$
and $\Delta_{1,N_E}^\alpha$ are nonzero.
The nonzero values of $\Omega_{i,j}^\alpha$ and $\Delta_{i,j}^\alpha$ are generated uniformly at random
from the range $[-4/3,4/3]$.

\begin{figure}[t]
\includegraphics[width=8cm]{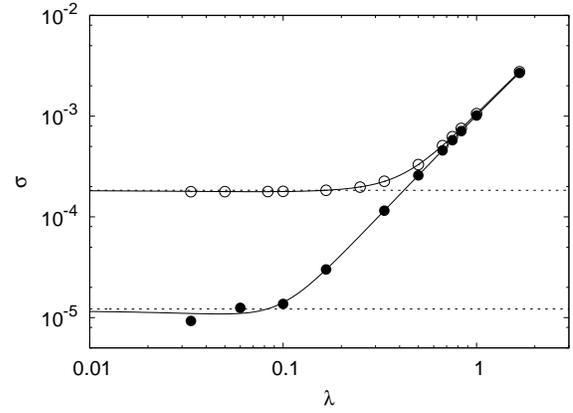}
\caption{Simulation results for $\sigma $
for a ring with $N_{S}=4$, $N_{E}=26$ (open circles) and $N_{S}=4$, $N_{E}=36$ (solid circles)
as a function of the global interaction strength $\lambda $ for $\beta |J|=6$.
For the values of the interaction parameters, see text.
The solid lines are fits to the data as described in the text.
The top (bottom) dashed line represents the value obtained by simulating the non-interaction system ($\lambda=0$)
with $30$ ($40$) spins.
}
\label{fig2}
\end{figure}

In Fig.~\ref{fig2} we present the simulation results for $\sigma$ as a function of $\lambda$.
Least square fitting of the data for $\sigma^2$ to polynomials in $\lambda$,
we find that a polynomial of degree $7$th yields the best fit, for both the 30- and 40-spin system data~\cite{30spins,40spins}.
The behavior of $\delta$ is very similar to that of $\sigma$ and is therefore not shown.
From Fig.~2 it is clear that for $\lambda\approx 1$,
the scaling of $\sigma$ with the dimension of the Hilbert space of the environment
is almost completely suppressed. This is a finite temperature effect:
for $\beta=0$, this scaling property holds independent of the coupling $\lambda$~\cite{JIN13a}.

Summarizing, in this Letter, we investigate the measures of decoherence
and thermalization of a quantum system $S$ coupled to
an environment $E$ at finite temperature. If the whole system $%
S+E$ is prepared in a canonical thermal state, we show
by means of perturbation theory that $\sigma^2$,
the degree of the decoherence of $S$, is of the order $\beta ^{2}\lambda ^{2}$. Up
to the first order in the system-environment interaction
we find
$\sigma^2 , \delta^2 \propto \exp\{-2\beta [F_{E}(2\beta )-F_{E}(\beta )]\}$
where $F_E$ is the free energy of the environment.
This provides a measure for how well a weakly-coupled specific finite environment can
decohere and thermalize a system at an inverse temperature $\beta $.
A measure for how difficult it is to decohere a quantum system is
given by ratios of free energies of the system, and at low temperatures
by $\sigma^2 \sim 1/2 g_E$ for a highly degenerate ground state for $S$
and a ground state degeneracy $g_E$ for $E$.
We performed numerical simulations of spin-$1/2$ systems to assess the region of validity
of the analytical results.

Strictly speaking, the system $S$ completely decoheres
if there is no interaction between $S$ and $E$ and if $N_E\rightarrow \infty$. If $S$
is coupled to $E$, the $S$--$E$ interaction is important.
Generally, $\sigma$ and $\delta$ for a finite system $S$
are finite even in the thermodynamic limit ($N_E\rightarrow\infty$). However, if the canonical
ensemble is a good approximation for the state of the system for some inverse temperatures $\beta$
up to some chosen maximum energy $E_{hold}>0$ (measured from the ground
state), then it is required that $\exp(-\beta E_{hold}) \gg \sigma$.
By determining the crossover of the left- and right-side
functions, we find a threshold for the temperature above which the state of
the system is well approximated by a canonical ensemble, and below which
quantum coherence of the system is well preserved.

\textsl{Acknowledgments}.
The authors gratefully acknowledge the computing time granted by the JARA-HPC
Vergabegremium and provided on the JARA-HPC Partition part of the supercomputer JUQUEEN at Forschungszentrum J{\"u}lich.
MAN is supported in part by US National Science Foundation grant DMR-1206233.

\bibliographystyle{apsrev4-1}
\bibliography{epr11,scaleH}

\begin{thebibliography}{34}
\bibitem [{30s()}]{30spins}%
  \BibitemOpen
  \href@noop {} {}\bibinfo {note} {{The coefficients of the $7$th-order
  polynomial fit to the data (see Fig.~2) of the 30-spin system are
  $a_0=0.34\times10^{-7}$, $a_1=-0.59\times10^{-7}$, $a_2=0.56\times10^{-6}$,
  $a_3=-0.18\times10^{-5}$, $a_4=0.41\times10^{-5}$, $a_5=-0.28\times10^{-5}$,
  $a_6=0.12\times10^{-5}$, and $a_7=-0.23\times10^{-6}$ }}\BibitemShut
  {NoStop}%
\bibitem [{40s()}]{40spins}%
  \BibitemOpen
  \href@noop {} {}\bibinfo {note} {{The coefficients of the $7$th-order
  polynomial fit to the data (see Fig.~2) of the 40-spin system are
  $a_0=0.14\times10^{-9}$, $a_1=-0.65\times10^{-9}$, $a_2=0.27\times10^{-8}$,
  $a_3=-0.12\times10^{-7}$, $a_4=0.11\times10^{-5}$, $a_5=0.46\times10^{-8}$,
  $a_6=-0.80\times10^{-7}$, and $a_7=0.12\times10^{-7}$ }}\BibitemShut
  {NoStop}%
\end{thebibliography}

\end{document}